\documentclass[sigconf, anonymous=False,review=False]{acmart}
\usepackage{multirow}
\usepackage{makecell}
\usepackage{subfig}
\usepackage{bbding}
\usepackage{enumitem}
\usepackage{colortbl}
\usepackage{diagbox}
\usepackage{comment}
\usepackage{amsmath}
\usepackage{algorithm}
\usepackage{algpseudocode}

\AtBeginDocument{%
  }
\usepackage{newfloat}
\usepackage{listings}

\setcopyright{acmlicensed}
\copyrightyear{2018}
\acmYear{2018}
\acmDOI{XXXXXXX.XXXXXXX}
\acmConference[Conference acronym 'XX]{Make sure to enter the correct
  conference title from your rights confirmation email}{June 03--05,
  2018}{Woodstock, NY}

\acmISBN{978-1-4503-XXXX-X/2018/06}

\begin{document}

\title{GeoGR: Enabling Spatio-Temporal Aware Industrial-scale Generative POI Recommendations} 
\author{Fangye Wang}
\authornote{Contributed equally to this research.}
\affiliation{%
  \institution{AMAP, Alibaba Group}
  \city{Beijing}
  \country{China}
}
\email{wangfangye.wfy@alibaba-inc.com}

\author{Haowen Lin}
\authornotemark[1]
\affiliation{%
  \institution{AMAP, Alibaba Group}
  \city{Beijing}
  \country{China}
}
\email{linhaowen.lhw@taobao.com}

\author{Yifang Yuan, Siyuan Wang}
\affiliation{%
  \institution{AMAP, Alibaba Group}
  \city{Beijing}
  \country{China}
}
\email{{yuanyifang.yyf,tuolei.wsy}@alibaba-inc.com}


\author{Xiaojiang Zhou}
\authornote{Corresponding Author}
\affiliation{%
  \institution{AMAP, Alibaba Group}
  \city{Beijing}
  \country{China}
}
\email{zhouxiaojiang.zxj@taobao.com}

\author{Song Yang}
\affiliation{%
  \institution{AMAP, Alibaba Group}
  \city{Beijing}
  \country{China}
}
\email{song.yangs@alibaba-inc.com}

\author{Pengjie Wang}
\affiliation{%
  \institution{AMAP, Alibaba Group}
  \city{Beijing}
  \country{China}
}
\email{pengjie.wpj@alibaba-inc.com}

\begin{abstract}
  Next Point-of-Interest (POI) prediction is a fundamental task in location-based services (LBS), especially critical for large-scale navigation platforms such as AMAP that serve billions of users in diverse lifestyle scenarios. Although recent POI recommendation approaches based on SIDs have achieved promising performance, they struggle in complex, sparse real-world environments due to two key limitations: (1) inadequate modeling of high-quality SIDs that capture cross-category spatio-temporal collaborative relationships, and (2) poor alignment between large language models (LLMs) and the POI recommendation task. To this end, we propose GeoGR, a geographic generative recommendation framework tailored for navigation-based LBS like AMAP, which perceives changes in users' contextual states and enables spatio-temporal aware POI recommendation. GeoGR features a two-stage design: (i) a geo-aware SID tokenization pipeline that explicitly learns spatio-temporal collaborative semantic representations via geographically constrained co-visited POI pairs, contrastive semantic representation learning, and iterative refinement; and (ii) a multi-stage LLM training strategy that aligns non-native SID tokens through continued pre-training with multiple prompt templates and enables autoregressive POI generation via supervised fine-tuning. Extensive experiments on multiple real-world datasets demonstrate GeoGR’s superiority over state-of-the-art baselines. Moreover, the deployment on the large-scale AMAP platform over three months, serving millions of users and delivering significant online gains of +2.91\% in WINRATE and +5.55\% in PV\_CTR, confirms its practical effectiveness and scalability in production. 
\end{abstract}



\maketitle
\section{Introduction}

Next Point-of-Interest (POI)~\cite{wang2025tool4poi,zhang2025survey,wang2025gnpr} prediction aims to infer the future destinations that a user is likely to visit based on historical trajectories and contextual signals such as time, location, and behavioral patterns. In large-scale navigation-centric Location-Based Service (LBS) platforms such as AMAP, this task serves as a core component of the search and recommendation system~\cite{wei2025oneloc, chen2025onesearch}, directly influencing query understanding, POI retrieval, and downstream user-facing ranking in the main search page~\cite{zhai2025cognitive}. Unlike prior academic studies that primarily rely on offline benchmarks, industrial POI recommendation systems must operate under stringent constraints of latency, scalability, and extreme data sparsity, while simultaneously supporting billions of users across highly heterogeneous scenarios, including navigation, dining, accommodation, tourism, and transportation services~\cite{zhang2025survey}.

Despite extensive progress in spatio-temporal recommendation and sequence modeling, existing approaches remain insufficient for real-world LBS deployment. Traditional methods formulate next-POI recommendation as a sequential recommendation problem and rely on manually designed architectures or embedding-based representations to capture user behavior patterns~\cite{kang2018self,song2024precise,sun2019rotate,jin2023language,yang2025sparse}. Recently, LLMs~\cite{yang2025qwen3, naveed2025comprehensive} have been increasingly adopted for their powerful reasoning and generalization capabilities, enabling more nuanced modeling of user intent by fusing historical trajectories with personal preferences and contextual cues~\cite{zhou2024plm4traj,yi2025recgpt,chen2024hllm,hong2025eager,zheng2024adapting}. However, most LLM-based approaches still represent POIs using either non-semantic identifiers or weak textual embeddings, which fail to capture rich collaborative relationships among POIs~\cite{wang2025gnpr, he2026plum}, particularly cross-category dependencies (e.g., airport–hotel–parking interactions) that naturally arise from real-world mobility behavior. Generative Recommendation (GR)~\cite{lv2026reasoning, zhai2025cognitive, wang2025gnpr,he2026plum} has emerged as a promising paradigm by rethinking POI tokenization. Rather than assigning each POI a unique atomic identifier, GR encodes it into a short sequence of discrete tokens—referred to as Semantic IDs (SIDs)—drawn from a shared, compact vocabulary~\cite{rajput2023recommender, ju2025generative}. Target POIs are then generated autoregressively, token by token, in a manner analogous to large language models (LLMs). While this paradigm offers a unified and scalable formulation for large-scale recommendation, most existing GR frameworks still suffer from two critical limitations when applied to industrial LBS scenarios.

First, a central limitation lies in the construction and utilization of SIDs. While SIDs provide a compact and generative-friendly representation of POIs, most existing approaches derive them purely from textual semantics, without explicitly modeling spatio-temporal collaborative signals such as geographic proximity, user mobility patterns, and functional complementarity. As a result, the learned SIDs fail to encode structurally meaningful POI relationships that are essential for location-aware recommendation. Second, existing methods integrate SIDs into LLMs by appending them as newly added tokens to the LLM vocabulary and directly optimizing them through standard supervised fine-tuning (SFT)~\cite{wei2021finetuned, wang2025gnpr}. However, since these newly introduced tokens are randomly initialized or only weakly aligned with the LLM's semantic space, they lack meaningful prior knowledge and do not naturally capture task-specific POI relationships. This creates a significant misalignment between general-purpose language modeling and domain-specific generation objectives, leading to unstable training and suboptimal performance in real-world production environments.

\begin{figure}[t]
    \centering
    \includegraphics[width=0.88\linewidth]{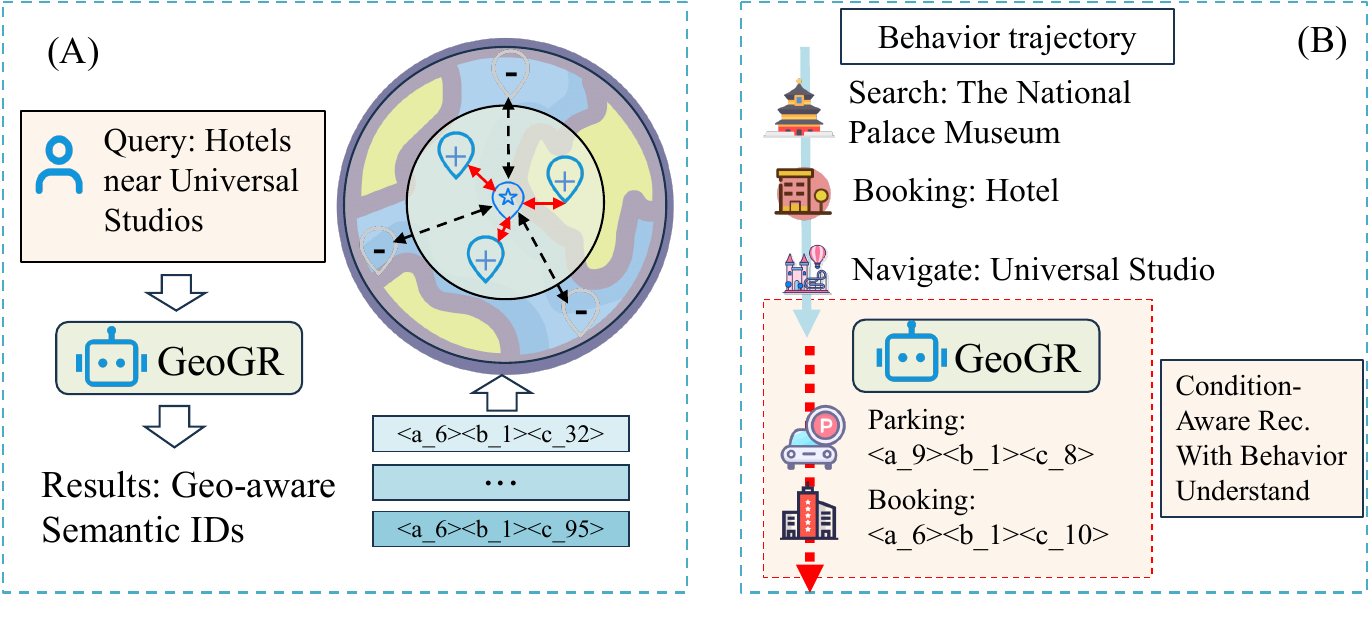}
    \caption{ Functions of GeoGR. (A) GeoGR generates geo-aware Semantic IDs from a user's query and spatial context. (B) GeoGR performs condition-aware recommendation by modeling the user's behavior trajectory. }
    \label{fig:intro_image}
\end{figure}

To address these two limitations, we propose GeoGR, a deployable generative recommendation framework integrated into the AMAP search system and serving the main POI recommendation pipeline. As illustrated in Figure \ref{fig:intro_image}, GeoGR supports two representative industrial scenarios: (A) query-driven POI retrieval, where the model generates geo-aware Semantic IDs conditioned on the user's explicit query and real-time location; and (B) behavior-sequence-based recommendation, where the model leverages historical trajectories and contextual signals for condition-aware next-POI recommendation. Training GeoGR consists of two core stages: SID construction and generative LLM training. In the SID construction stage, we design a three-step pipeline: (i) learning spatio-temporal collaborative semantic representations via contrastive learning on geographically constrained co-visitation pairs, (ii) discretizing representations into hierarchical SIDs using RQ-Kmeans over LLM embeddings, and (iii) refining SIDs with an EM-style iterative optimization procedure. In the generative stage, we perform CPT using a mixture of text and SID hybrid data with two objectives: aligning the new SID token with base LLM and injecting scenario-specific recommendation knowledge, followed by SFT using user behavior data to generate concrete POI recommendations conditioned on user contextual information, e.g., profile, location, time, and query. GeoGR has been deployed in the AMAP platform and serves millions of daily active users in production. Online A/B experiments demonstrate substantial and consistent gains over the production baseline, including +2.91\% WINRATE and +5.55\% PV\_CTR, confirming its practical effectiveness in large-scale LBS scenarios.
 
The major contributions are summarized as follows:
\begin{itemize}[leftmargin=1em]
    \item  We propose GeoGR, an industrial generative recommendation framework integrated into the AMAP search system for large-scale next-POI recommendation in real-world LBS scenarios.
    
    \item We design a geo-aware SID tokenization pipeline that constructs compact and stable SIDs in three steps: learning spatio-temporal collaborative semantic representations via contrastive learning; generating discrete SIDs using RQ-Kmeans; refining SIDs with an EM-style iterative optimization algorithm.
    
    \item We introduce a multi-stage alignment and fine-tuning strategy to adapt LLMs to the next POI recommendation task, including CPT-based template alignment for non-native SID tokens and SFT for autoregressive POI generation.
    
    \item Extensive offline and online A/B experiments on AMAP prove significant and consistent improvements over strong baselines, confirming the superiority and industrial applicability of GeoGR. In addition, we will open-source the industrial datasets.
\end{itemize}

\section{Related Works}
\textbf{POI Recommendation.}  Traditional next-POI recommendation is typically formulated as a sequential recommendation task, where the goal is to predict a user’s next visited location based on their historical check-in sequence~\cite{zhai2025cognitive,li2024llm4poi_rec,wang2025gnpr,kang2018self,sun2019bert4rec}. Early approaches leverage Markov chains or matrix factorization to model transition patterns or user–POI affinities but struggle to capture long-term dependencies and complex spatio-temporal dynamics. With the advent of deep learning, recurrent neural networks (RNNs)—particularly LSTM and GRU variants—were widely adopted to model long-term user preferences and sequential dependencies~\cite{hidasi2015session}. Subsequently, transformer-based models, powered by self-attention mechanisms, demonstrated superior performance over RNNs in next-POI recommendation~\cite{kang2018self}. For instance, STAN~\cite{luo2021stan} employs self-attention to explicitly model pairwise spatio-temporal relationships among POIs, capturing both non-consecutive transitions and geographic proximity through interpolated distance encoding. Rotan~\cite{feng2024rotan} encodes time intervals through rotational position vector representations in transformer architectures, effectively capturing temporal dynamics in user behavior sequences. More recently, the integration of large language models (LLMs) has opened new frontiers in semantic-aware location modeling. LLM4POI~\cite{li2024llm4poi_rec} fine-tunes LLMs on standard-scale datasets to leverage commonsense knowledge for next-POI recommendation. 


\textbf{Semantic IDs and Generative Recommendation.}  
RecSys performance hinges on learning quality representations~\cite{cheng2016wide, wang2023gdcn}. The dominant paradigm assigns unique but uninformative IDs to users and items, learning embeddings that primarily encode collaborative signals~\cite{loveland2025role, rendle2020neural}. While effective in dense settings, this approach struggles with scalability and with sparse or long-tail items, where co-occurrence signals are insufficient~\cite{ju2024does, ju2025generative}. Semantic IDs (SIDs) address these limitations by first encoding rich semantic features (e.g., textual descriptions) through modality encoders such as LLMs, then quantizing the resulting dense embeddings into compact, discrete token sequences. Common quantization-based tokenizers include RQ-VAE~\cite{lee2022autoregressive}, RVQ~\cite{van2017neural}, and RQ-Kmeans~\cite{deng2025onerec}. Because SIDs are derived from semantic content rather than randomly assigned, they naturally generalize to unseen or rare items and reduce vocabulary sparsity~\cite{ju2025generative}. This compact token representation has given rise to Generative Recommendation (GR)~\cite{liu2024end,wei2025oneloc, zhang2025survey}, which rethinks the recommendation pipeline as an autoregressive generation task: instead of scoring every candidate item, the model directly generates the SID of the target item token by token. TIGER~\cite{rajput2023recommender} pioneers this two-stage framework by learning hierarchical SIDs and then generating candidates via autoregressive decoding. In industrial settings, KuaiShou's OneRec~\cite{deng2025onerec} and OneRec-V2~\cite{zhou2025onerec} further unify retrieval and ranking into an end-to-end generative framework with reinforcement learning. Furthermore, GNPR-SID~\cite{wang2025gnpr} introduces a generative next-POI recommendation model that integrates SIDs to jointly capture intrinsic POI semantics, collaborative signals, and user visit patterns. In the local lifestyle services domain, KuaiShou further develops OneLoc~\cite{wei2025oneloc}, an end-to-end generative recommender that fully leverages geographic information.

\section{Preliminaries}
In this section, we present the key notation and formulate the next-POI recommendation problem. Let $\mathcal{U} = \{u_1, u_2, ..., u_M\}$ and $\mathcal{P} = \{p_1, p_2, ..., p_N\}$ denote the sets of $M$ users and $N$ POIs, respectively. Each POI $p_i \in \mathcal{P}$ is described by geographic coordinates, address, and contextual attributes such as category, brand, and name. For each user $u \in \mathcal{U}$, the historical interaction trajectory is denoted as $\mathcal{T}^u = \{r^u_1, r^u_2, ..., r^u_n\}$, where each record $r^u_i = \langle u, p, t, con, a \rangle$ indicates that user $u$ interacted with POI $p$ at time $t$ under condition $con$ (e.g., time, location, weather) with action $a$ (e.g., click, navigate, book). The records are arranged in reverse chronological order.

\textbf{Next POI Recommendation.} Given a user $u_i$, the interaction history $\mathcal{T}^{u_i}$, and a real-time contextual condition  $con_u$, the task is to predict the next POI $p_{n+1} \in \mathcal{P}$ with which the user will interact. This is formulated as maximizing the conditional probability:
\begin{align}
\mathscr{p}(p_{n+1} \mid u_i, \mathcal{T}^{u_i}, con_u).
\end{align}
In the generative recommendation paradigm, the model directly generates the target POI's SID $\mathbf{Q}_{p_{n+1}}$ autoregressively:
\begin{equation}
    \begin{aligned}
\mathscr{p}(\mathbf{Q}_{p_{n+1}} \mid u_i, \mathcal{T}^{u_i}, con_u; \theta) = \prod_{j=1}^{L} \mathscr{p}(q^{(j)}_{p_{n+1}} \mid u_i, \mathcal{T}^{u_i}, con_u, q^{(<j)}_{p_{n+1}}; \theta)
    \end{aligned}
\end{equation}
where $L$ is the number of SID layers and $\theta$ denotes the model parameters. This generative formulation unifies retrieval and ranking into a single autoregressive decoding process, which is especially suitable for large-scale industrial deployment.

\section{Methods: GeoGR}
\begin{figure*}
    \centering
    \subfloat[Construct the Geo-aware POI Semantic ID with three steps: (1) learning spatio-temporal collaborative semantic representation with contrastive learning; (2) Generating the POI SID with RQ-Kmeans cluster; (3) Improving the POI SID with EM-style SID optimization algorithm. ]{
     \begin{minipage}[t]{0.95\linewidth}
     \centering
     \includegraphics[width=0.80\textwidth]{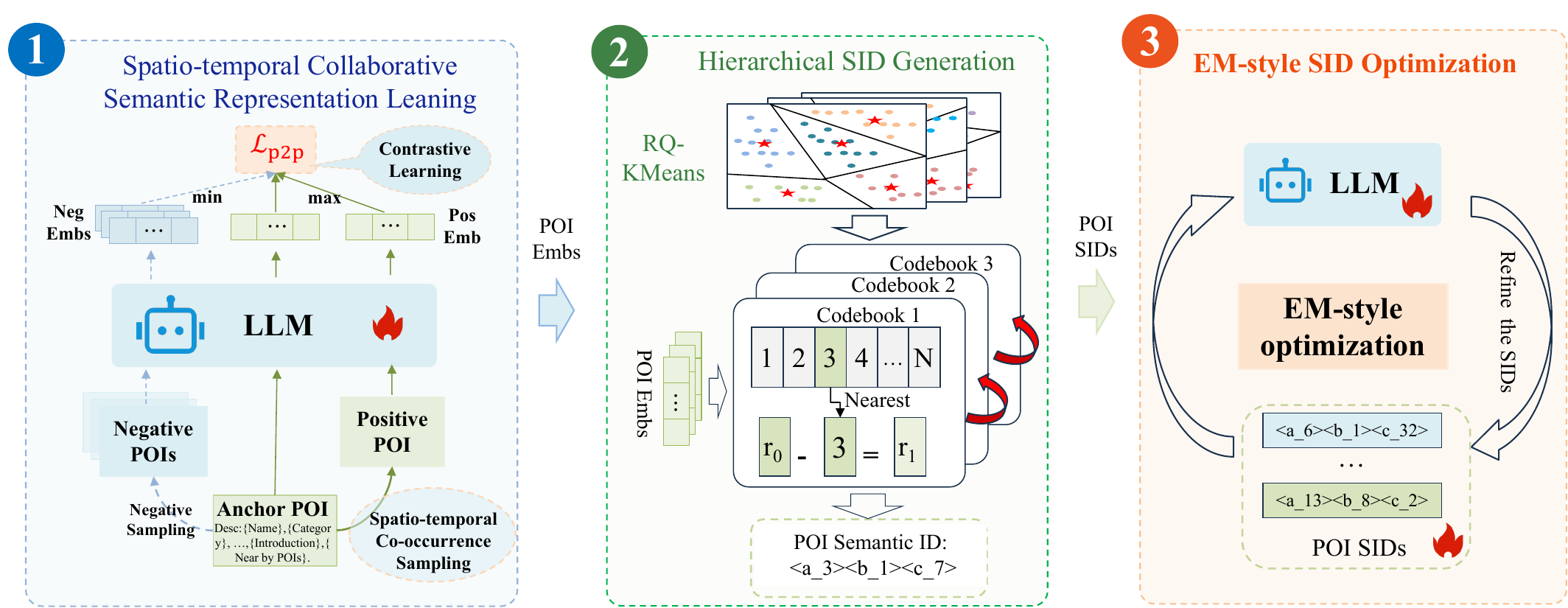}
     \end{minipage}
    }
    \vspace{-1em}
    
    \subfloat[Train the generative POI recommendation model with two steps: (1) Continued pre-training; (2) Supervised fine-tuning.]{
     \begin{minipage}[t]{0.99\linewidth}
     \centering
     \includegraphics[width=0.85\textwidth]{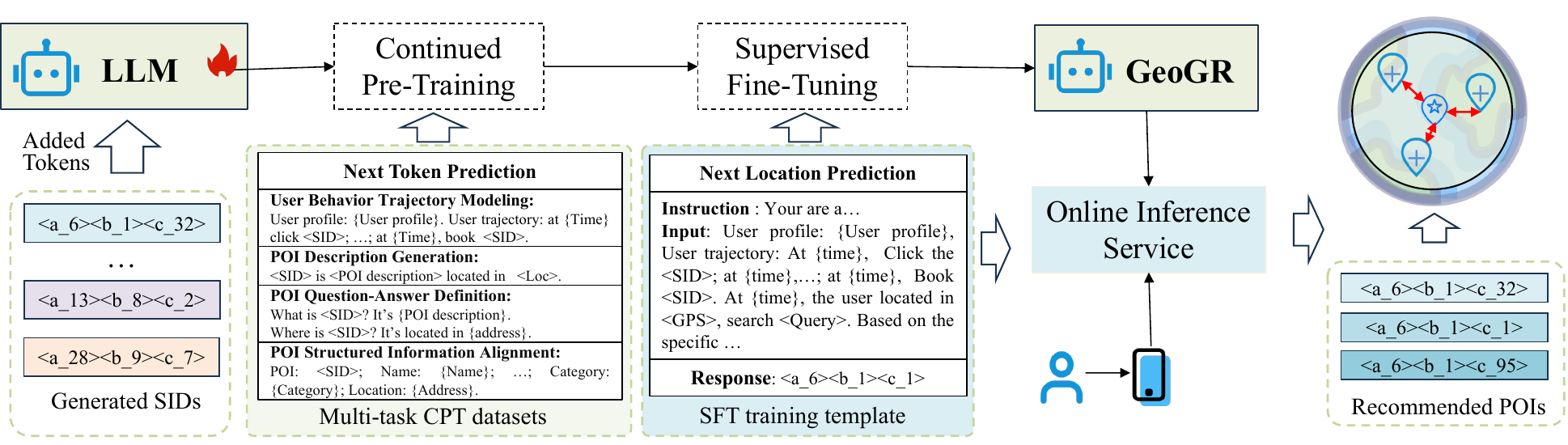}
     \end{minipage}
    }
    \caption{The overall architecture of GeoGR.}
\label{fig:GeoGR}
\end{figure*}

As illustrated in Figure ~\ref{fig:GeoGR}, GeoGR follows a two-stage pipeline. Its core innovation lies in encoding the rich semantics of POIs into compact, discrete tokens that seamlessly integrate with LLMs, thereby enhancing their ability to understand user intent based on contextual conditions, search queries, and mobility trajectories.

In the first stage, we encode textual and geographic information for each POI—including longitude, latitude, location address, and contextual descriptors (e.g., name, brand, category)—into a set of specialized location SIDs using hierarchical vector quantization techniques such as RQ-Kmeans. The resulting SIDs preserve rich semantic information while ensuring that semantically similar POIs are mapped to nearby representations through the hierarchical codebook structure. More importantly, a compact set of codebook tokens can efficiently represent a vast number of POIs, circumventing the vocabulary explosion problem that arises when introducing excessive new tokens into LLMs. Furthermore, the short length and well-defined structure of the SIDs improve the fine-tuning stability, leading to a more accurate POI generation.

In the second stage, we train the LLM to progressively understand the recommendation task and ultimately perform condition-aware retrieval and prediction of the next POI. First, we conduct CPT to align the newly introduced SID tokens with POI and recommendation corpus over multiple template-based corpora. Building upon this CPT-LLM, we construct an instruction-tuning dataset specifically designed for next-location prediction and further refine the model via supervised fine-tuning. Through this process, GeoGR achieves a deep understanding of user behavior and delivers high-quality next-POI recommendations.

\subsection{Geo-aware POI SID Construction } 
\label{sec:sid}

Most existing LLM-based POI recommendation methods rely on randomly assigned identifiers (RIDs), which lack semantic meaning and fail to capture user intent in contextual scenarios. Moreover, such random encoding drastically increases model storage requirements, hindering scalability and inference efficiency.
To address these limitations, we adopt an RQ-Kmeans based semantic tokenization strategy~\cite{deng2025onerec} that leverages POI textual descriptions to learn a coarse-to-fine hierarchical codebook through hierarchical vector quantization. By using a reduced, fixed-size set of tokens, our approach alleviates storage and computational overhead while enabling seamless integration of new POIs—each mapped to semantically similar existing tokens—thereby improving generalization. Given the stringent requirement for geographic precision in navigation scenarios, our tokenization strategy comprises three key steps: (1) learning spatio-temporal collaborative semantic-aware POI representations, (2) semantic representation tokenization, and (3) SID optimization via an EM-style algorithm \cite{dempster1977maximum}. These steps collectively enhance the spatio-temporal awareness of POIs, embed rich semantic information into SID tokens, and iteratively refine the SIDs through EM-style optimization guided by LLM. 

\subsubsection{Spatio-temporal Collaborative Semantic Representation}
\label{sec:p2p}
Here, we aim to map each POI description into a POI representation vector for subsequent SID generation. To fully encode both semantic and geographic information, we construct an informative location description $p_i= \{lon, lat, loc, c \}$, which includes geographical coordinates (longitude and latitude), location address, and contextual attributes such as name, brand, categories, geohash code, consumption level, active time period. Existing approaches typically feed POI textual descriptions into a pre-trained LLM to obtain semantic embeddings. While semantically similar POIs tend to yield close representations—and local context can help capture spatial proximity—such purely semantic encoding suffers from two key limitations: (1) it overlooks POI pairs that exhibit strong spatiotemporal associations despite weak semantic similarity; (2) it dilutes the encoding of the target POI’s intrinsic semantics by attending to overly long sequences of neighboring POIs. To address these issues, we augment POI representation learning with a contrastive fine-tuning strategy that explicitly models pairwise geographic distances. Specifically, as shown in Figure~\ref{fig:GeoGR}(a), we adopt three steps:

\textbf{POI Semantic Representation.} Here, we map the basic POI textual descriptions into an off-the-shelf LLM (i.e., Qwen~\cite{yang2025qwen3}) to obtain semantic embeddings:
\begin{align}
    \mathbf{e}_{p_i} = LLM_{encoder}(concat(lon,lat, loc, c)) .
    \label{equ:llm_encoder}
\end{align}

\textbf{Geo-Constrained Co-visited POI Pairs.} In the above textual POI descriptions, to emphasize the intrinsic attributes of each POI, we refrain from directly incorporating neighboring POIs or collaborative signals from multiple users into the input text. Instead, we explicitly model the joint constraints of geographic proximity and behavioral co-occurrence by constructing spatiotemporal co-visit POI pairs. Specifically, we first sample POI pairs from user trajectories that frequently co-occur in sequences—even if semantically dissimilar—using Swing similarity to downweight popular items and highlight genuine behavioral affinity. We then filter these pairs by geographic distance (e.g., 3 km), retaining only those that exhibit both strong collaborative signals and spatial proximity.

\textbf{POI-to-POI Contrastive Loss.} We fine-tune the aforementioned LLM using a Noise-Contrastive Estimation (NCE) loss over the constructed spatiotemporal co-visit POI pairs. For each positive pair  $ (\mathbf{e}_{p_i}, \mathbf{e}_{p_j}) $ , the contrastive loss is defined as:
\begin{align}
    \mathcal{L}_{\text{P2P}}(i, j) = -\log \frac{\exp(\mathbf{e}_{p_i}^\top \mathbf{e}_{p_j} / \tau)}{\exp(\mathbf{e}_{p_i}^\top \mathbf{e}_{p_j} / \tau) + \sum_{k \in \mathbb{N}^-} \exp(\mathbf{e}_{p_i}^\top \mathbf{e}_{p_k} / \tau)}.
\end{align}
where  $\mathbb{N}^- $  denotes the set of negative POIs and  $ \tau > 0 $  is the temperature coefficient.  Using the fine-tuned encoder  $ LLM_{\text{encoder}} $ , we then generate the POI semantic representation following Equation~\ref{equ:llm_encoder}.

\subsubsection{Semantic Representation Quantization} 
Based on the spatially aware semantic representations, we employ RQ-Kmeans for tokenization, which leverages residual quantization to transform POI embeddings into coarse-to-fine SIDs through a hierarchical set of discrete codebooks. The RQ-Kmeans procedure is illustrated in Figure~\ref{fig:GeoGR}(a). The initial residual set $\mathcal{R}^{(1)}=\{\mathbf{e}^1_i|i=1,2,...,N \}$ consists of the representations of all POIs, and each representation $\mathbf{e}^1_i$ is generated by the fine-tuned LLM in previous step. 

For each layer $l$, we apply RQ-Kmeans clustering on the residual set $\mathcal{R}^{(l)}$ to compute the codebook $\mathcal{C}^{(l)}$: 
\begin{align}
\textstyle
\mathcal{C}^{(l)} = Kmeans(\mathcal{R}^{(l)}, N_c)
\end{align}
where $\mathcal{C}^{(l)} = \{ \mathbf{c}^{(l)}_k | k=1,2,...,N_c \} $ denotes the set of centroids obtained from K-means, and $N_c$ is the codebook size. Using $\mathcal{C}^{(l)}$, we can compute the nearest centroid index $q^l_i$ for each residual vector as:
\begin{align}
q^{(l)}_i= \arg\min_k ||\mathcal{R}^{(l)}_i - \mathbf{c}^{(l)}_k ||,
\end{align}
where $||\cdot||$ is the Euclidean distance.  Index $q^{(l)}_i$ corresponds to the $l$-th level SID token. The residual for the next layer is updated as:
\begin{align}
\textstyle
\mathcal{R}^{(l+1)}_{i} = \mathcal{R}^{(l)}_{i} - \mathbf{c}^{(l)}_{q^{(l)}_i}.
\end{align}

In this work, we repeat the quantization process for  $ L_c = 3 $  layers, yielding three codebooks  $ (\mathcal{C}^{(1)}, \mathcal{C}^{(2)}, \mathcal{C}^{(3)}) $ . Consequently, each POI with descriptive input  $ \mathbf{p}_i $  is mapped to a hierarchical SID:
\begin{align}
    \mathbf{Q}_{p_i} = \text{Tokenizer}(\mathbf{e}_{p_i}) = [q^{(1)}_i, q^{(2)}_i, q^{(3)}_i].
\end{align}
The full set of SIDs across all POIs is denoted as  $ \mathbb{P} = \{ \mathbf{Q}_{p_i} \mid i = 1, 2, \dots, N \} $ . We represent a three-level SID as  $ \langle a\_5 \rangle \langle b\_3 \rangle \langle c\_8 \rangle $ , where the tokens correspond to centroid indices  $ \mathbf{c}^{(1)}_5 $ ,  $ \mathbf{c}^{(2)}_3 $ , and  $ \mathbf{c}^{(3)}_8 $  from the first, second, and third codebook layers, respectively. Empirically, this hierarchical structure enables efficient progressive POI generation in the recommendation system.

\begin{algorithm}[t]
\caption{EM-style SID optimization algorithm}
\label{alg:em}
\begin{algorithmic}[1]
\Require Base LLM $LLM_{\text{ori}}$; initial SIDs set $\mathbb{P}$; POI info set $\mathbf{P}$
\Ensure Optimized SIDs $\mathbb{P} \gets \mathbb{P}_n$

\State $\mathbb{P}_0 \gets \mathbb{P}$ 
\State $LLM_{\text{0}} \gets LLM_{\text{ori}}$ \Comment{e.g., Qwen-4B}

\For{$i = 1$ to $n$}
    \State Construct SFT training dataset using POI descriptions $\mathbf{P}$ and current SIDs $\mathbb{P}$;
    \State Fine-tune $LLM_{\text{i-1}}$ on the SFT dataset to obtain $LLM_i$;
    \State Predict new SIDs: $\mathbb{P}_{\text{i}} \gets LLM_i(\mathbf{P})$; 
    \State Compute $\text{Acc}_i$: quantile accuracy between $\mathbb{P}_{\text{i}}$ and $\mathbb{P}_{i-1}$;
    \State Compute $\text{HitRate}_i$: evaluate $LLM_i$ on downstream task;
\EndFor

\State \Return $\mathbb{P}_n$ \Comment{Final optimized SIDs $\mathbb{P}_n$}
\end{algorithmic}
\end{algorithm}

\subsubsection{SID Optimization with EM-style Algorithm}
\label{sec:em}

The quality of SIDs critically affects the effectiveness of downstream SFT of large language models. To improve SID quality, we propose an EM-style SID optimization algorithm that iteratively refines SIDs through self-improvement with an LLM (e.g., Qwen-4B). As shown in Algorithm~\ref{alg:em}, the framework alternates between two phases: (1) fixing the SIDs and fine-tuning the LLM to better predict them, and (2) fixing the LLM and updating the SIDs based on its predictions. Our core insight is that, since the ultimate goal is to generate SIDs using the LLM in the next-POI recommendation task, we should leverage its predictive capability from the outset and use its outputs to refine the initial SIDs—thereby encouraging mutual alignment between the LLM and the SIDs. Specifically, during the SID prediction step (e.g., Line 6 in Algorithm~\ref{alg:em}), we employ dynamic beam search to generate a ranked list of the 20 candidate SIDs to prevent SID collisions. If the ground-truth target SID (i.e., the true next POI) is not included in this list, we replace it with the highest-probability SID from the generated candidates, ensuring that the final output always corresponds to a high-confidence prediction.

\subsection{Generative POI Recommendation}

Building upon the learned POI SIDs, we now describe the generative model training module of GeoGR. We adopt an instruction-tuning paradigm to adapt a pre-trained LLM backbone (Qwen-4B) to the next-POI recommendation task. Since the newly introduced SID tokens are out-of-vocabulary for the base LLM, we first perform Continued Pre-Training (CPT) to warm up these tokens and inject scenario-specific recommendation knowledge. This CPT stage aligns the SID vocabulary with the LLM's semantic space and enables the model to capture POI-related mobility patterns. We then apply SFT on cross-scenario user behavior sequences, optimizing the model for next-POI recommendation under real-world contextual conditions. The SFT training instructions are kept consistent with the online inference template to ensure deployment fidelity.

\subsubsection{CPT Optimization}

Before performing SFT on user behavior sequences, we prepare multi-task CPT datasets with various prompt templates, which are shown in Figure \ref{fig:GeoGR}(b). By jointly optimizing over these heterogeneous contexts using large-scale unsupervised data, CPT achieves two key objectives: (1) Warming up SID tokens: The model learns to treat added SID tokens as grounded semantic entities rather than abstract symbols;  (2) Alignment of scenario-specific recommendation knowledge: Through various prompt templates, we inject multiple types of POI recommendation–related knowledge, such as POI information and user behavior sequences—into LLM. This alignment substantially enhances downstream task performance by preserving the expressiveness and personalization of natural language while ensuring structured, controllable generation through SID-based outputs. Formally, the CPT process is framed as a next-token prediction task. Given a token sequence  $ s = (s_1, \dots, s_T) $  drawn from a corpus that interleaves SID tokens with associated real-world POI attributes and user interaction contexts, the base LLM with parameters  $ \theta_0 $  is trained to maximize the log-likelihood of the observed sequence under the autoregressive assumption:
\begin{align}
    \mathcal{L}_{\text{CPT}}({\theta_0}) = \sum\nolimits_{t=1}^{T} \log \mathscr{p}_{\theta_0}(s_t \mid s_{<t}).
\end{align}
After training, we can obtain a CPT-LLM with parameters $ \theta_1  $. 

\subsubsection{SFT Optimization}

Here, we adopt next-POI recommendation as the primary fine-tuning objective, consistent with established approaches in the literature. Using comprehensive, cross-scenario POI and user behavior data, we construct a high-quality instruction-tuning dataset to train the large language model (LLM) to predict the most likely next-location Semantic ID (SID).  Specifically, each training example follows an instruction-based prompt format comprising three components:  (1) an \textit{Instruction} that explicitly specifies the prediction task;  (2) an \textit{Input} containing the user’s current context (e.g., location, time, search query) and short-term behavioral trajectory; and   (3) a \textit{Response} consisting of the target SID.  Formally, the structured prompt is defined as follows: 
\begin{center}
    \includegraphics[width=0.99\linewidth]{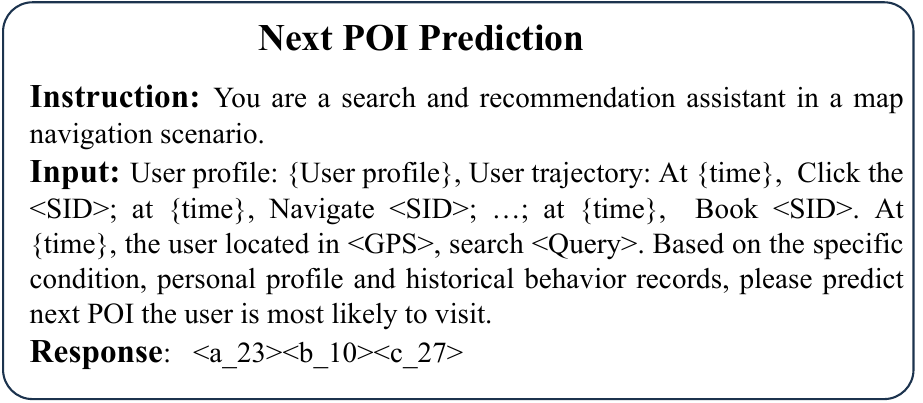}
\end{center}

The full prompt comprises both the instruction and input components. The short-term user trajectory consists of the most recent  $ L_s = 32 $  interacted POIs (as used in our experiments). The current context includes the query issuance location, search query, timestamp, and weather conditions. Additionally, we construct a personalized user profile by summarizing the user’s long-term historical trajectories, capturing their activity preferences along with temporal and spatial visitation patterns and associated frequencies. By using compact, summarized representations instead of raw, ultra-long trajectories, we achieve higher training efficiency without compromising recommendation quality. This design enables the LLM to jointly encode and reason over both short- and long-term behavioral signals, thereby improving prediction performance. Meanwhile, geo-aware SIDs ensure the model remains sensitive to the spatial characteristics of POIs, facilitating accurate and context-aware next-POI recommendations.

We perform full-parameter SFT based on the CPT-trained LLM. Given the above prompt structure, the SFT objective is to minimize the negative log-likelihood (NLL) loss:
\begin{equation}
    \begin{aligned}
        \mathcal{L}_{\text{NLL}}(\theta_1) 
        &= -\log \mathscr{p}_{\theta_1}(\mathbf{Q}_t \mid \text{prompt})  \\ 
        &= -\sum_{i=1}^{3} \log \mathscr{p}_{\theta_1}\big( q^{(i)} \mid \text{prompt}, q^{(<i)} \big),
    \end{aligned} 
\end{equation}
where  $ \mathbf{Q}_t = [q^{(1)}, q^{(2)}, q^{(3)}] $  denotes the SID of the target POI.

Based on the above steps, we obtain the SFT-LLM as the POI generative recommender (GR), i.e., GeoGR. During inference, GeoGR generates the most likely SID tokens autoregressively given the input prompt, which corresponds to approximately maximizing the conditional probability $\mathscr{p}_{\theta_2}(\mathbf{Q}_t \mid \text{prompt}) = \prod_{i=1}^{3} \mathscr{p}_{\theta_2}\big(q^{(i)} \mid \text{prompt}, q^{(<i)}\big), $  where  $ \theta_2 $  denotes the GeoGR's parameters. In the candidate generation phase, we employ beam search (or alternative decoding strategies) to obtain the top-$k$  most probable SIDs according to  $\mathscr{p}_{\theta_2}(\mathbf{Q}_t \mid \text{prompt})$.

\section{Experiments}
\subsection{Experimental Setting}
\subsubsection{Datasets.}  
We conduct experiments on four real-world datasets: two public Foursquare check-in datasets (NYC and TKY) and two large-scale industrial datasets from the AMAP platform (AMAP and AMAP-SH). Detailed statistics are provided in Table~\ref{tab:dataset}.

\textbf{Public datasets.} NYC and TKY~\cite{yang2014modeling,wang2025gnpr} contain user POI check-in trajectories collected over an 11-month period in New York City and Tokyo, respectively. Following the preprocessing protocol of GNPR-SID~\cite{wang2025gnpr}, we filter out POIs with fewer than 10 interactions and users with fewer than 10 check-ins. The records are temporally sorted and split into 80\% training, 10\% validation, and 10\% testing.  \textbf{Industrial datasets.} The AMAP and AMAP-SH datasets are constructed from real user interactions on the AMAP platform\footnote{\url{https://amap.com}}. Both datasets capture user intent through search queries and rich contextual signals, including GPS coordinates and target-location metadata, with the goal of retrieving the most relevant POIs for display on the app's homepage. For the AMAP dataset, we process one day of platform-wide traffic. For each active user on that day, we trace back their historical interaction records to construct a complete behavior sequence, used as a training instance. For evaluation, we randomly sample 1 million records from the following day's interactions as the test set. AMAP-SH is a subset sampled from Shanghai over a one-month period. Interactions from the last day are held out as the test set, and others are used for training.

\subsubsection{Evaluation Metrics.}  
Following prior work~\cite{wei2025oneloc,wang2025gnpr}, we adopt two widely used metrics to evaluate recommendation performance: top-$K$ Recall (Recall@$K$) and Normalized Discounted Cumulative Gain (NDCG@$K$). We report both metrics with $K \in \{5, 10, 20\}$ to assess recommendation performance at varying retrieval depths. Higher values of both metrics indicate better recommendation quality. Meanwhile, the \textit{Rela.Imp} (Relative Improvement) quantifies performance gains relative to a base model.

\begin{table}[t]
    \centering
    \caption{Statistics of the processed datasets.}
    \setlength{\tabcolsep}{3.1mm}
    \scalebox{0.95}{
    \begin{tabular}{ccccc}
    \hline
    \hline
         Dataset & \#Users & \#Items &\#Records  & Sparsity  \\
         \hline
         NYC &2,083 &5,135 &104,074  &98.10\% \\ 
         TKY &2,293 &7,873 &361,430  & 98.98\%\\
         \hline
         AMAP&7,536K&34,119K &2,298,665K&99.99\% \\ 
         AMAP-SH&4,820K& 910K& 22,333K  & 99.99\% \\ 
    \hline
    \hline
    \end{tabular}
    }
    \label{tab:dataset}
\end{table}

\begin{table*}[t]
    \setlength{\abovecaptionskip}{0.2cm}
    \setlength{\belowcaptionskip}{-0.2cm}
\centering
\caption{Offline performance on NYC and TKY. "Rela.Imp" indicates the relative improvement compared to the best baseline.}
\label{exp:overall}
\setlength{\tabcolsep}{2.5mm}
\scalebox{0.75}{
    \begin{tabular}{c|c|ccccc|cc|ccc|ccc} 
    \hline
    \hline
    \multicolumn{1}{c|}{\multirow{2}{*}{Dataset}} & \multirow{2}{*}{Metrics} & \multicolumn{5}{c|}{Traditional} & \multicolumn{2}{c|}{POI Related} & \multicolumn{3}{c|}{LLM-based}& \multicolumn{1}{c}{\multirow{2}{*}{\makecell{GeoGR\\-0.6B}}} & \multicolumn{1}{c}{\multirow{2}{*}{\makecell{GeoGR\\-4B}}} &\multicolumn{1}{c}{\multirow{2}{*}{Rela.Imp}}  \\ 
    \cline{3-12}
    \multicolumn{1}{c|}{}  &  & \multicolumn{1}{l}{SASRec} & \multicolumn{1}{c}{BERT4Rec} & \multicolumn{1}{l}{GRU4Rec} & \multicolumn{1}{c}{Caser} & \multicolumn{1}{c|}{$S^3$-Rec} & \multicolumn{1}{l}{TPG} & \multicolumn{1}{c|}{Rotan} & \multicolumn{1}{c}{TIGER} & \multicolumn{1}{l}{GNPR-SID} & \multicolumn{1}{c|}{OneLoc} & \multicolumn{1}{c}{} & \multicolumn{1}{c}{} & \multicolumn{1}{c}{} \\ 
    \hline
    \multirow{6}{*}{NYC}   
     & Recall@5 & 0.3151 & 0.2857& 0.1977  & 0.2883& 0.3071& 0.3551   & 0.4448 & 0.4965& 0.5311&  \underline{0.6107}  &  \textit{0.6214} &\textbf{0.6446} &	\textbf{5.55\%} \\
     & Recall@10& 0.3896 & 0.3564& 0.2460  & 0.3570& 0.3854& 0.4441   & 0.5223 & 0.5514& 0.5942& \underline{0.6563}   &\textit{0.6604} &\textbf{0.6897} &	\textbf{5.09\%} \\
     & Recall@20& 0.4506 & 0.4130& 0.2889  & 0.4135& 0.4503& 0.5121   & 0.5834 & 0.6001& 0.6455& \underline{0.6977}  &\textit{0.7041} &\textbf{0.7349} &	\textbf{5.33\%} \\
     \cline{2-15}
     & NDCG@5   & 0.2224 & 0.2074& 0.1442  & 0.2044& 0.2235& 0.2464   & 0.3471 & 0.4131& 0.4430& \underline{0.5355}  &\textit{0.5373} &\textbf{0.5484} &	\textbf{2.41\%} \\
     & NDCG@10  & 0.2467 & 0.2304& 0.1599  & 0.2267& 0.2489& 0.2755   & 0.3723 & 0.4276& 0.4634& \underline{0.5504}  &\textit{0.5518}&\textbf{0.5621} &	\textbf{2.13\%} \\
     & NDCG@20  & 0.2622 & 0.2448& 0.1708  & 0.2410& 0.2654& 0.2927   & 0.3878 & 0.4443& 0.4766& \underline{0.5608}  &\textit{0.5636}&\textbf{0.5730} &	\textbf{2.18\%} \\
    \hline
    \multirow{6}{*}{TKY}   
     & Recall@5 & 0.3450 & 0.2649& 0.2514  & 0.3257& 0.3365& 0.3725   & 0.4333 & 0.5031& 0.5354& \underline{0.5964}  &\textit{0.6027} &\textbf{0.6259} &	\textbf{4.95\%} \\
     & Recall@10& 0.4284 & 0.3326& 0.3106  & 0.4067& 0.4115& 0.4601   & 0.5113 & 0.5808& 0.6130& \underline{0.6620}  &\textit{0.6710}&\textbf{0.6908} &	\textbf{4.35\%} \\
     & Recall@20& 0.4976 & 0.3943& 0.3651  & 0.4758& 0.4739& 0.5291   & 0.5894 & 0.6431& 0.6675& \underline{0.7152}  &\textit{0.7163}&\textbf{0.7382} &	\textbf{3.22\%} \\
     \cline{2-15}
     & NDCG@5   & 0.2384 & 0.1907& 0.1833  & 0.2273& 0.2423& 0.2591   & 0.3293 & 0.4003& 0.4437& \underline{0.4961}  &\textit{0.5025}&\textbf{0.5127} &	\textbf{3.35\%} \\
     & NDCG@10  & 0.2655 & 0.2127& 0.2025  & 0.2535& 0.2666& 0.2881   & 0.3568 & 0.4251& 0.4623& \underline{0.5174}  &\textit{0.5252}&\textbf{0.5311} &	\textbf{2.65\%} \\
     & NDCG@20  & 0.2831 & 0.2284& 0.2163  & 0.2711& 0.2825& 0.3051   & 0.3739 & 0.4401& 0.4788& \underline{0.5306}  & \textit{0.5319}&\textbf{0.5423} &	\textbf{2.21\%} \\
    \hline
    \hline
    \end{tabular}
}
\end{table*}


\begin{table*}[t]
    \centering
    \caption{Overall performance on large-scale dataset.}
    \scalebox{0.80}{
    \label{exp:amap_all}
    \begin{tabular}{c|ccc|ccc|ccc|ccc}
    \hline
    \hline
    \multicolumn{1}{c|}{Dataset}  & \multicolumn{6}{c|}{AMAP} & \multicolumn{6}{c}{AMAP-SH} \\
    \hline
    Method & Recall@5 & Recall@10 & Recall@20 & NDCG@5 & NDCG@10 & NDCG@20 & Recall@5 & Recall@10 & Recall@20 & NDCG@5 & NDCG@10 & NDCG@20 \\
    \hline
    TIGER    & 0.3815 & 0.4326 & 0.4935 & 0.2883 & 0.3142 & 0.3360 &0.3426	&0.4283	&0.5040	&0.2328	&0.2763	&0.3007\\
    GNPR-SID & 0.4306 & 0.4921 & 0.5836 & 0.3005 & 0.3338 & 0.3564 &0.3593	&0.4416	&0.5394	&0.2501	&0.2894	&0.3214\\
    OneLoc   & \underline{0.4543} & \underline{0.5382} & \underline{0.6114} & \underline{0.3419} & \underline{0.3794} & \underline{0.3983} &\underline{0.3915}	&\underline{0.4791}	&\underline{0.5716}&	\underline{0.2830}	&\underline{0.3249}	&\underline{0.3531}\\
    \hline
    GeoGR-4B & \textbf{0.5068} & \textbf{0.6138} & \textbf{0.7059} & \textbf{0.3836} &  
    \textbf{0.4184} & \textbf{0.4418} &\textbf{0.4527}	&\textbf{0.5384}&\textbf{0.6437}	&\textbf{0.3292}	&\textbf{0.3731}	&\textbf{0.4089}\\
    \textbf{Rela.Imp} & \textbf{11.56\%} & \textbf{14.05\%} & \textbf{15.46\%} & \textbf{12.20\%} & \textbf{10.28\%} & \textbf{10.92\%} &\textbf{15.63\%}	&\textbf{12.38\%}	&\textbf{12.61\%}	&\textbf{16.33\%}	&\textbf{14.84\%}	&\textbf{15.80\%} \\
    \hline
    \hline
    \end{tabular}
    }
\end{table*}

\subsubsection{Comparison Methods.}  We compare GeoGR with three categories of baselines: (1) \textit{Traditional sequential methods}: SASRec~\cite{kang2018self}, BERT4Rec~\cite{sun2019bert4rec}, GRU4Rec~\cite{hidasi2015session}, Caser~\cite{tang2018caser},  $S^3$-Rec~\cite{zhou2020s3};  (2) \textit{Transformer-based POI-related methods}: TPG~\cite{luo2023timestamps}, Rotan~\cite{feng2024rotan};   (3) \textit{Generative LLM-based methods}: TIGER~\cite{rajput2023recommender}, GNPR-SID~\cite{wang2025gnpr}, OneLoc~\cite{wei2025oneloc}. Detailed introduction of baselines can be found in the referenced papers.

\subsubsection{Implementation Details.}  All experiments are implemented
in PyTorch~\cite{paszke2019pytorch}. Specifically, the quantization module employs three codebook layers. We configure the codebook sizes according to dataset scale: for NYC and TKY, the sizes are set to 32 and 64, respectively; for the industrial-scale datasets AMAP and AMAP-SH, the size is set to 4096 to accommodate the substantially larger POI vocabulary. We adopt Qwen-4B~\cite{yang2025qwen3} consistently across all components of GeoGR: as the POI semantic embedding model, as the backbone for EM-style SID optimization, and as the base LLM for CPT and full-parameter fine-tuning in the generative recommendation module. We use a constant learning rate of $1 \times 10^{-5}$ with a warm-up phase of 20 steps. Training is conducted on 16 H20 GPUs with a per-GPU batch size of 16. The gradient accumulation steps are set to 2 or 4 and adjusted according to memory constraints. 

\subsection{Main Results}

\subsubsection{Comparison of Public Datasets}
Table~\ref{exp:overall} reports the performance on two public benchmarks, NYC and TKY. We make the following observations:

First, SID-based generative LLM-based methods consistently outperform traditional sequential models and POI-specific approaches across all metrics. This demonstrates that conventional recommenders—relying on rigid embeddings and manually designed architectures—struggle to capture nuanced user mobility intent. In contrast, LLM-based methods leverage rich semantic signals for comprehensive intent understanding and deep contextual reasoning, yielding substantially superior recommendation performance.

Second, GeoGR achieves the best performance across all settings. GeoGR-4B consistently outperforms the strongest baseline, while GeoGR-0.6B, despite its significantly smaller parameter count, also surpasses all traditional, POI-specific, and generative baselines. Specifically, on the NYC dataset, GeoGR-4B improves over the best baseline by 5.55\%, 5.09\%, and 5.33\% in Recall@5/10/20, and by 2.41\%, 2.13\%, and 2.18\% in NDCG@5/10/20, respectively. Consistent improvements are observed on TKY, confirming the robustness and generalizability of our approach.

\subsubsection{Comparison of Industrial Datasets}
To further validate GeoGR's scalability and effectiveness in real-world production environments, we evaluate it on two large-scale industrial datasets collected from the AMAP platform: AMAP, which covers nationwide user behaviors, and AMAP-SH, a subset focused on Shanghai. Table~\ref{exp:amap_all} reports the offline performance.

Compared with the public benchmarks (NYC and TKY), the industrial datasets present substantially greater challenges due to their massive POI vocabularies, extreme data sparsity, and highly heterogeneous user behaviors across cities and scenarios. Despite these challenges, GeoGR-4B consistently outperforms all baselines by a much larger margin than on the public datasets. On AMAP, GeoGR-4B improves over the strongest baseline OneLoc by 11.56\%, 14.05\%, and 15.46\% in Recall@5/10/20, and by 12.20\%, 10.28\%, and 10.92\% in NDCG@5/10/20. The gains are even more pronounced on AMAP-SH, with relative improvements of 15.63\%, 12.38\%, and 12.61\% in Recall@5/10/20, and 16.33\%, 14.84\%, and 15.80\% in NDCG@5/10/20. Averaged across all metrics, GeoGR surpasses OneLoc by approximately 12.4\% on AMAP and 14.6\% on AMAP-SH, compared to 3.22\%-5.55\% on Recall and 2.13\%-3.35\% on NDCG for NYC/TKY.


%

This significant performance gap highlights two key advantages of GeoGR. First, the geo-aware SID construction becomes increasingly critical as the dataset scale grows: industrial POIs exhibit complex spatio-temporal collaborative relationships that cannot be captured by textual semantics alone. By explicitly modeling geographically constrained co-visitation patterns, GeoGR learns more discriminative POI representations than OneLoc and other baselines. Second, the CPT+SFT multi-stage alignment effectively bridges the gap between general LLM semantics and industrial recommendation objectives. While baselines such as TIGER and GNPR-SID also employ SIDs, they lack dedicated alignment mechanisms for non-native SID tokens, resulting in weaker generalization under sparse and noisy industrial data. Overall, these results demonstrate that GeoGR not only achieves strong performance on standard academic benchmarks but also delivers substantial practical gains in large-scale LBS production environments.

\begin{table}[t] 
\centering
\caption{Ablation study on AMAP dataset.}
\label{exp:abl}
 \scalebox{0.65}{
    \begin{tabular}{c|ccc|ccc}  
    \hline
    \hline
    Variants  & Recall@5 & Recall@10 & Recall@20 & NDCG@5 & NDCG@10 & NDCG@20\\
    \hline
    BaseGR & 0.3440&	0.4145	&0.4766	&0.2631	&0.2860	&0.3018 \\
    \hline 
    \multirow{2}{*}{with $\mathcal{L}_{P2P}$}
    &0.4472	&0.5508&	0.6234	&0.3514	&0.3785	&0.3969 \\
    &($\uparrow$30.00\%)	&$(\uparrow$32.88\%)	&($\uparrow$30.80\%)	&($\uparrow$33.56\%)	&($\uparrow$32.34\%)	&$(\uparrow$31.51\%) \\
    \hline 
    \multirow{2}{*}{with EM}
    &0.4004		&0.4834		&0.5372		&0.3283		&0.3540		&0.3768	 \\
    &($\uparrow$16.40\%)		&($\uparrow$16.62\%)		&($\uparrow$12.72\%)		&($\uparrow$24.78\%)		&($\uparrow$23.78\%)		&($\uparrow$24.85\%)   \\
    \hline
    \multirow{2}{*}{with $\mathcal{L}_{P2P}$\&EM}
    &0.4732		&0.5676		&0.6493		&0.3617		&0.3923		&0.4131	\\
    &($\uparrow$37.56\%)		&($\uparrow$36.94\%)		&($\uparrow$36.24\%)		&($\uparrow$37.48\%)		&($\uparrow$37.17\%)	&($\uparrow$36.88\%)   \\
    \hline
    Full
    & \textbf{0.5068}	&\textbf{0.6138}	&\textbf{0.7059}	& \textbf{0.3836}    &\textbf{0.4184}	&\textbf{0.4418} \\
    ($\mathcal{L}_{P2P}$\&EM\&CPT) &($\uparrow$47.33\%)		&($\uparrow$48.08\%)		&($\uparrow$48.11\%)		&($\uparrow$45.80\%)		&($\uparrow$46.29\%)		&($\uparrow$46.39\%)  \\
    \hline
    \hline
    \end{tabular}
}
\end{table} 

\subsection{Ablation Study} 
We conduct an ablation study to verify the contribution of each key component in GeoGR. We consider the following variants: (1) \textbf{BaseGR} constructs SIDs purely from POI textual semantics (i.e., Base SID) via RQ-Kmeans and fine-tunes the LLM with standard SFT; (2) \textbf{$\mathcal{L}_{P2P}$} additionally learns spatio-temporal collaborative semantics through geographically constrained POI-to-POI contrastive learning; (3) \textbf{EM} applies EM-style iterative optimization to refine SIDs; (4) \textbf{$\mathcal{L}_{P2P}$ \& EM}, which combines both SID improvements; and (5) \textbf{Full ($\mathcal{L}_{P2P}$ \& EM \& CPT)} further adds the CPT stage on top of the previous variant. The results are reported in Table~\ref{exp:abl}.

\textbf{Effectiveness of geo-aware SID construction.} Both the spatio-temporal collaborative representation learning ($\mathcal{L}_{P2P}$) and the EM-style SID optimization improve SID quality. Adding $\mathcal{L}_{P2P}$ alone improves Recall@5 and NDCG@5 by 30.00\% and 33.56\% over BaseGR, respectively, demonstrating the importance of modeling geographic co-visitation patterns. EM alone also yields consistent gains of 16.40\% and 24.78\% on the same metrics, confirming that iterative refinement aligns SIDs with the generative model. Combining both modules further improves Recall@5 and NDCG@5 by 37.56\% and 37.48\%, indicating their complementary effects: $\mathcal{L}_{P2P}$ injects geographic and collaborative priors into embeddings, while EM preserves these structures through refined token assignments.

\textbf{Effectiveness of CPT.} The CPT stage provides an additional significant boost. Compared with the $\mathcal{L}_{P2P}$ \& EM variant, the Full model improves Recall@5 from 0.4732 to 0.5068 and NDCG@5 from 0.3617 to 0.3836, corresponding to relative gains of approximately 7.10\% and 6.05\%, respectively. This validates that aligning newly introduced SID tokens with the base LLM through CPT is crucial for stabilizing generative recommendation and injecting scenario-specific knowledge before SFT. Overall, the progressive improvements from BaseGR $\rightarrow$ $\mathcal{L}_{P2P}$ $\rightarrow$ EM $\rightarrow$ $\mathcal{L}_{P2P}$ \& EM $\rightarrow$ Full confirm that each component of GeoGR contributes meaningfully to the final performance.

\begin{table}[t]
    \setlength{\abovecaptionskip}{0.2cm}
    \setlength{\belowcaptionskip}{-0.4cm}
\centering
\caption{Evaluation of various SIDs in AMAP Dataset.   }
\label{exp:sim}
\scalebox{0.89}{
\begin{tabular}{c|ccc|ccc|c} 
\hline
    \hline
    Metrics & \multicolumn{3}{c|}{Semantic Similarity $\uparrow$} & \multicolumn{3}{c|}{Location Dis.(KM) $\downarrow$} & \multirow{2}{*}{Recall@5}   \\ 
    \cline{1-7}
   Cluster & C1 &C2  & C3 & C1 & C2 & C3&\\ 
\hline
    Base SID  & 0.62  & 0.73  & 0.90  & 483.4  & 71.2  & 11.2 & 0.3618 \\
    GNPR-SID  & 0.71  & 0.88  & 0.93  & 195.1  & 24.7  & 6.2 &0.4574  \\
    Our SID & \textbf{0.82}  & \textbf{0.98}  & \textbf{0.99}  & \textbf{31.9}   & \textbf{4.8}   & \textbf{1.3}  & \textbf{0.5068}\\
\hline
    \hline
\end{tabular}
}
\end{table}

\subsection{Semantic and Distance Relevance of SID} 

We evaluate the semantic similarity and real-world geographic distance among POIs (on the AMAP dataset) based on their assigned SID codebooks—namely, C1, C2, and C3. We compute the average semantic similarity and the average geographic distance among all POIs sharing the same prefix token (e.g., $<a_1, *, *>$).

The results are shown in Table~\ref{exp:sim}. Base SID, generated by basic Qwen-4B embeddings and RQ-Kmeans, exhibits weak geographic coherence: POIs sharing the same C1 prefix are on average 483.4 km apart, indicating that textual semantics alone fail to ground POIs in real-world geography. GNPR-SID improves over Base SID by incorporating temporal and collaborative signals as well as GeoHash information, achieving higher semantic similarity and substantially reducing geographic distance. Our SID further pushes both metrics toward optimal alignment through spatio-temporal collaborative signals and EM-style refinement, reaching 0.82/0.98/0.99 semantic similarity and 31.9/4.8/1.3 km geographic distance. This progressive improvement in SID quality directly corresponds to the gains in Recall@5: from 0.3618 (Base SID) to 0.4574 (GNPR-SID) and 0.5068 (Our SID). These results validate our core motivation for designing hierarchical SIDs: POIs that are semantically related and geographically close should be mapped to nearby or identical SIDs, and such alignment is critical for accurate next-POI recommendation.

\begin{table}[t]
    \centering
    \caption{Impact of CPT across various SID.}
    \label{tab:sid_ablation}
    \setlength{\tabcolsep}{3.8mm}
    \scalebox{0.75}{
    \begin{tabular}{cc|cc|c}
        \hline
        \hline
        \textbf{Metric} & \textbf{SID Method} & \textbf{Only SFT} & \textbf{CPT\&SFT} & \textbf{Rela.Imp} \\
        \hline
        \multirow{4}{*}{Recall@5}   
                 & Base SID & 0.3440 & \textbf{0.3618} & \textbf{5.17\%} \\
                 & GNPR SID & 0.4306 & \textbf{0.4574} & \textbf{6.22\%} \\
                 & Our SID  & 0.4732 & \textbf{0.5068} & \textbf{7.10\%} \\
                          \cline{2-5}
         & \textbf{Ave.Imp}  & --     & --  & \textbf{6.17\%}  \\
        
        \hline
        \multirow{4}{*}{NDCG@5}  
           & Base SID & 0.2631 & \textbf{0.2771} & \textbf{5.32\%} \\
                 & GNPR SID & 0.2969 & \textbf{0.3205} & \textbf{7.95\%} \\
                 & Our SID  & 0.3617 & \textbf{0.3836} & \textbf{6.05\%} \\
                 \cline{2-5}
         & \textbf{Ave.Imp}  & --     & -- & \textbf{6.44\%}  \\
        \hline
        \hline
\end{tabular}
}
\end{table}

\subsection{Impact of CPT across Various SID Methods}

To verify whether CPT generalizes across different SID construction strategies, we evaluate it on the three SID methods described above: Base SID, GNPR-SID, and Our SID. The results are shown in Table~\ref{tab:sid_ablation}. 



Applying CPT consistently and substantially improves recommendation performance across all three SID methods, demonstrating its strong effectiveness and broad compatibility. Averaged over Base SID, GNPR-SID, and Our SID, CPT improves Recall@5 and NDCG@5 by 6.17\% and 6.44\%, respectively. These universal gains indicate that CPT is a \textbf{plug-and-play alignment mechanism} for non-native SID tokens, which does not depend on a specific tokenization pipeline and can be reliably applied even when SID construction strategies vary or evolve across industrial scenarios. Furthermore, under the SFT-only setting, Our SID already substantially outperforms the other two SIDs, and combining Our SID with CPT achieves the strongest overall performance. These results confirm that CPT and high-quality geo-aware SIDs are complementary rather than substitutive, jointly forming a synergistic foundation for accurate next-POI recommendation.




\begin{figure}[t]
    \setlength{\abovecaptionskip}{0.2cm}
    \setlength{\belowcaptionskip}{-0.4cm}
    \centering
        \subfloat[Recall w.r.t. model size]{
         \begin{minipage}[t]{0.49\linewidth}
         \centering
         \includegraphics[width=0.99\textwidth]{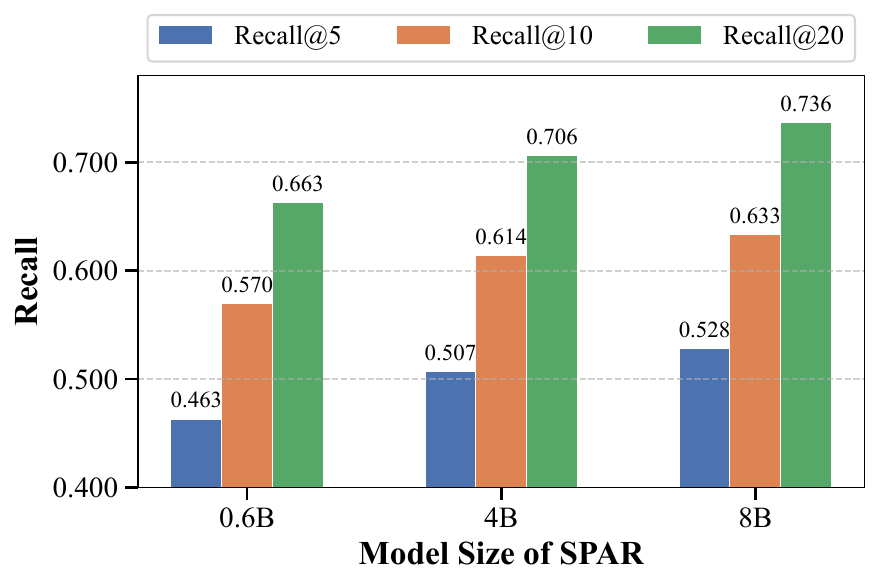}
         \end{minipage}
        }
        \subfloat[NDCG w.r.t. model size]{
         \begin{minipage}[t]{0.49\linewidth}
         \centering
         \includegraphics[width=0.99\textwidth]{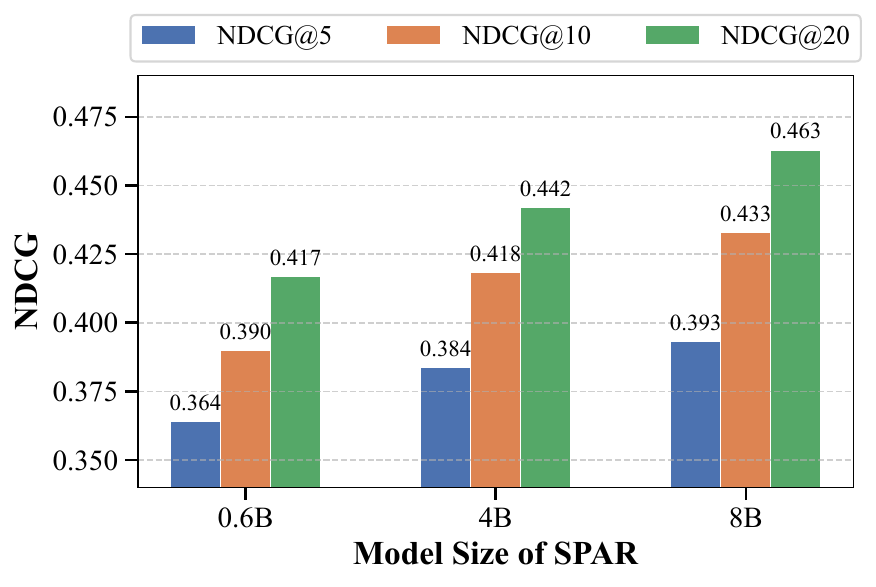}
         \end{minipage}
        }
        \caption{Impact of Model Size on GeoGR Performance. }
\label{exp:scaling}
\end{figure}

\subsection{Impact of Model Size}
To investigate how backbone model capacity affects retrieval performance, we evaluate GeoGR variants with 0.6B, 4B, and 8B parameters. The results are shown in Figure \ref{exp:scaling}.

As model size increases, both Recall@K and NDCG@K improve consistently. Scaling from GeoGR-0.6B to GeoGR-8B yields gains of 14.0\%, 11.2\%, and 11.1\% on Recall@5/10/20, and 8.0\%, 11.0\%, and 10.3\% on NDCG@5/10/20, respectively. These results indicate that larger language model backbones provide stronger representation capacity for geo-aware POI retrieval. However, considering the strict low-latency requirements of the AMAP homepage retrieval service, we deploy GeoGR-0.6B for long-term online serving.

\begin{figure}[t]
    \setlength{\abovecaptionskip}{0.2cm}
    \setlength{\belowcaptionskip}{-0.4cm}
    \centering
        \subfloat[Before, basic representation]{
         \begin{minipage}[t]{0.49\linewidth}
         \centering
         \includegraphics[width=0.85\textwidth]{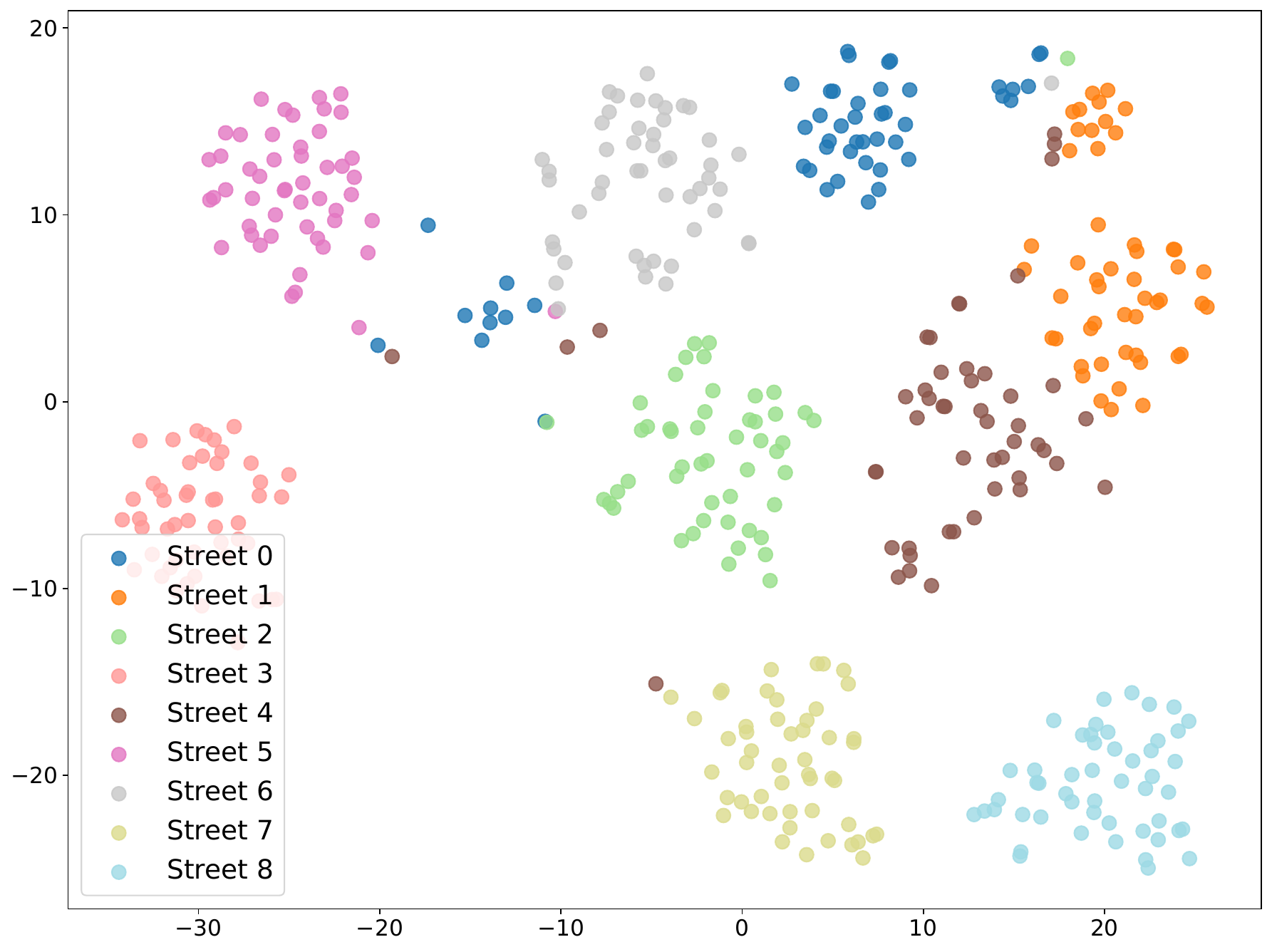}
         \end{minipage}
        }
        \subfloat[After, geo-aware representation]{
         \begin{minipage}[t]{0.49\linewidth}
         \centering
         \includegraphics[width=0.85\textwidth]{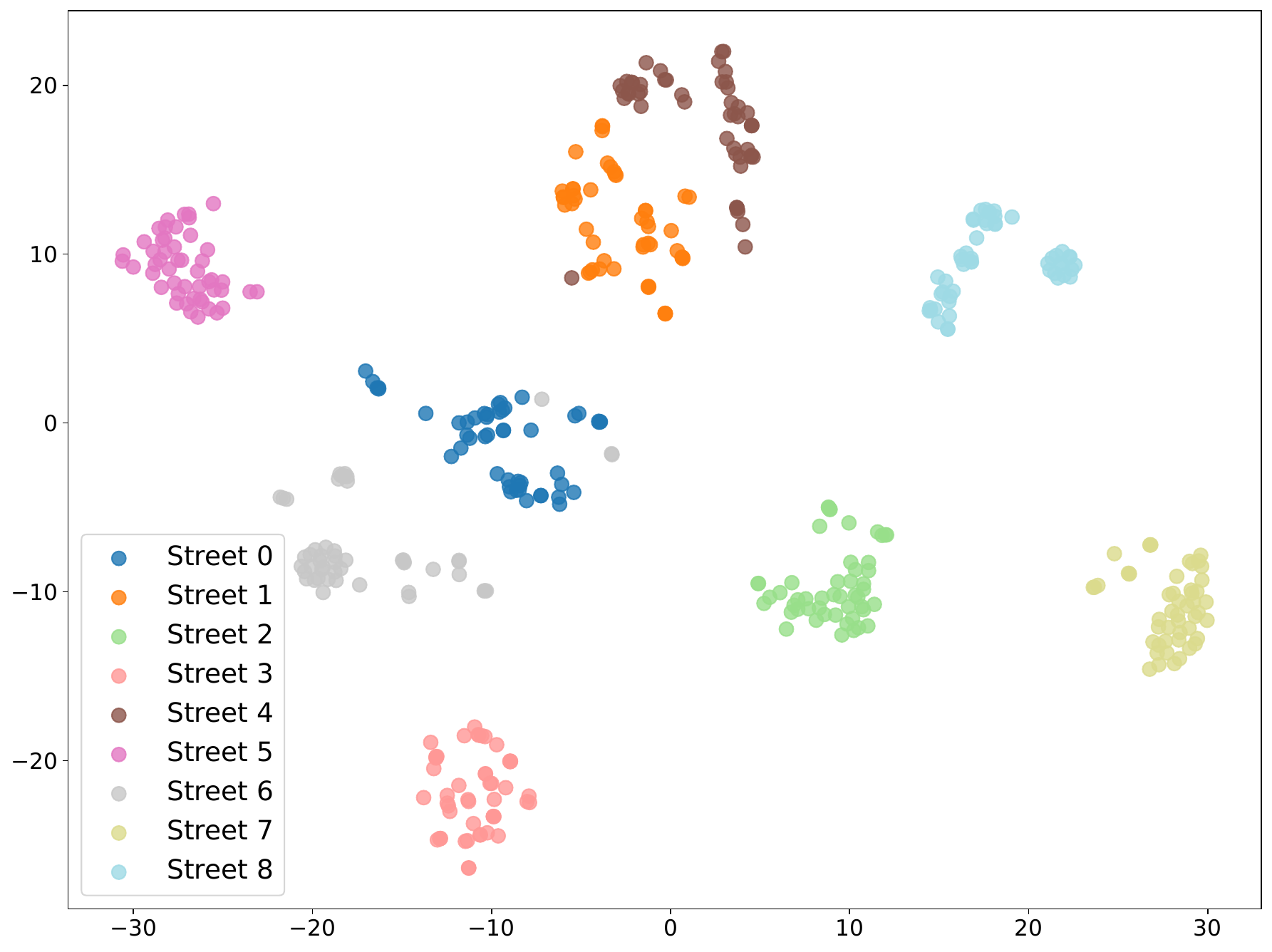}
         \end{minipage}
        }
        \caption{Visualization of POI representations before and after applying the spatio-temporal collaborative optimization. }
\label{exp:sid_vis}
\end{figure}

\subsection{Visualization Analysis}  

To intuitively illustrate the effect of spatio-temporal collaborative representation learning, we visualize POI embeddings using t-SNE~\cite{maaten2008visualizing} before and after optimization. We sample 450 POIs from nine streets in a city (50 POIs per street) and color them by street.

As shown in Figure~\ref{exp:sid_vis}, POIs from different streets are much better separated after applying geo-aware optimization. Before optimization, the basic representations learned from raw textual attributes exhibit noticeable entanglement across streets, suggesting that native LLM embeddings alone cannot fully capture fine-grained spatial semantics even when address and coordinate information is provided. After optimization, POIs on the same street cluster more tightly, while different streets form distinct groups. This aligns with the intuition that POIs on the same street are geographically close and more likely to co-occur in local user mobility trajectories. The visualization results corroborate the quantitative gains observed in the ablation studies, confirming that constructing geo-constrained co-visited POI pairs and optimizing representations through contrastive learning effectively enhances the geo-awareness of POI embeddings.

\begin{figure}[t]
    \centering
    \includegraphics[width=0.99\linewidth]{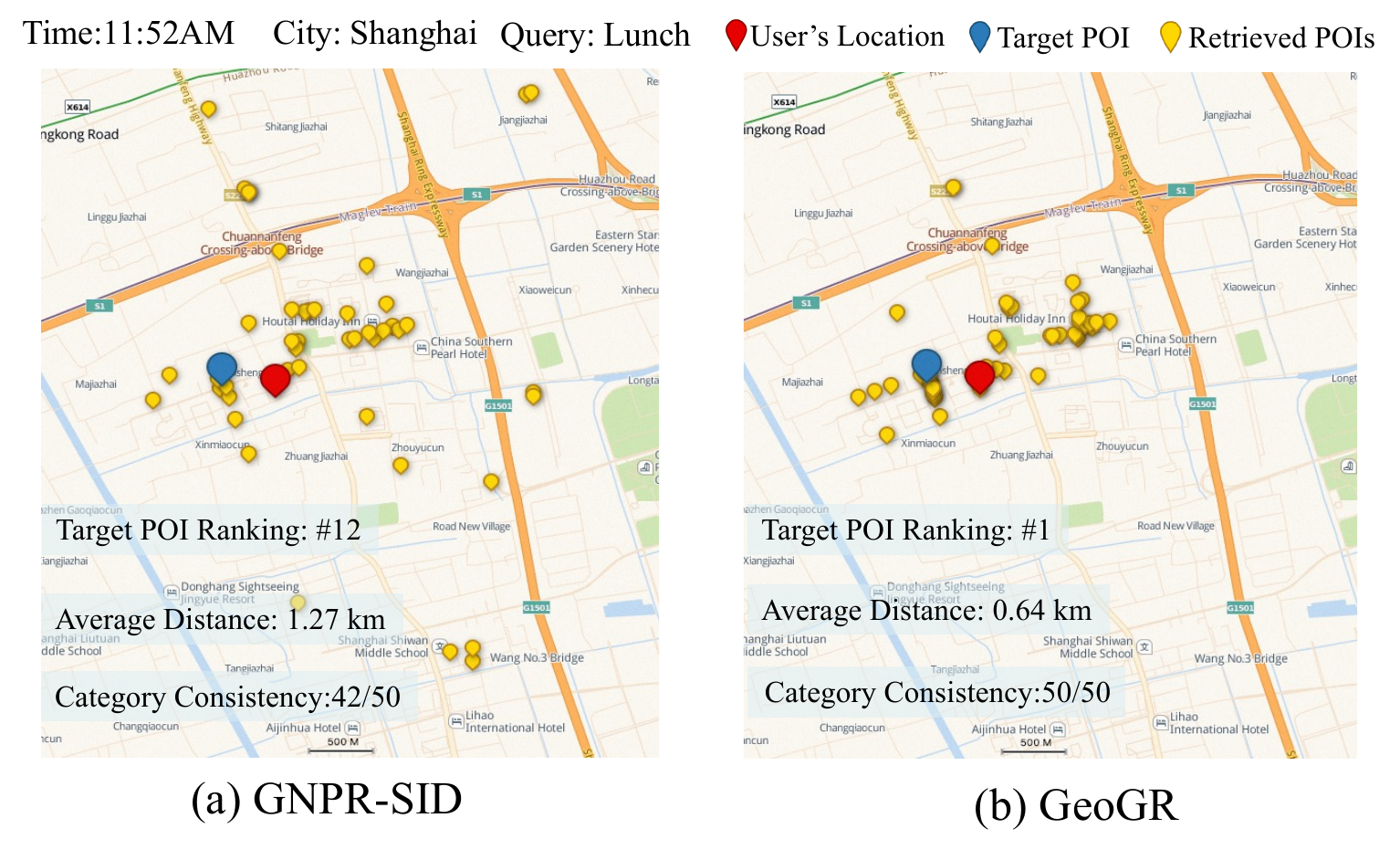}
    \caption{ Visualization of recommended POIs.}
    \label{fig:vis_demo}
\end{figure}

\subsection{Case Study}

Figure~\ref{fig:vis_demo} presents a qualitative case study comparing GNPR-SID and GeoGR for a user in Shanghai searching "Food" at 11:52AM. The red, blue, and yellow pins mark the user's location, the target POI, and the top-50 candidate POIs, respectively. We measure retrieval quality by Target POI Ranking, Average Distance to the user/target, and Category Consistency with the category of the Target POI.

As shown in Figure \ref{fig:vis_demo}(a), GNPR-SID identifies a non-trivial number of semantically relevant POIs, achieving a category consistency of 42/50. Nevertheless, its retrieved candidates are spatially dispersed: the average distance reaches 1.27 km and the target POI is ranked only \#12. Many candidates are spread across distant roads and even across major traffic corridors, suggesting that GNPR-SID prioritizes semantic relevance over geographic proximity without adequately conditioning on the user's immediate location and real-time context. In contrast, GeoGR (Figure \ref{fig:vis_demo}(b)) produces a markedly more concentrated retrieval set around both the user and the target. It reduces the average distance to 0.64 km—a 49.6\% decrease—and ranks the target POI at \#1, while simultaneously improving category consistency to 50/50. These gains stem from GeoGR's explicit spatio-temporal grounding: geo-aware SID construction pulls nearby POIs together in the embedding space, explaining the shorter average distance; CPT+SFT training aligns SIDs with real POI semantics, lifting category consistency from 42/50 to 50/50; and condition-aware generation exploits real-time context to rank the true target at \#1. GNPR-SID lacks such grounding, so its candidates scatter geographically even when semantically plausible.

\subsection{Online Experiments}

GeoGR has been deployed on the AMAP platform to power user query retrieval on the homepage, replacing the conventional multi-stage recall and coarse-ranking pipeline. When a user submits a query, the prompt construction module retrieves user-specific data from the database—including personalized preferences and historical behavior trajectories—and combines it with real-time contextual signals to form a complete input prompt. GeoGR then generates 50 candidate SIDs using dynamic beam search. After necessary post-processing, these candidates are passed to downstream fine-ranking and re-ranking modules to produce the final recommendations. To balance recommendation quality and inference efficiency, we adopt GeoGR-0.6B as the production model. It achieves an average latency of 50 ms per request on NVIDIA H20 GPUs, compared to 90 ms for GeoGR-4B. Although GeoGR-4B demonstrates stronger offline performance, its higher latency makes it less suitable for long-term production deployment.

We have conducted a large-scale online A/B test of GeoGR-0.6B against the legacy multi-stage retrieval baseline for three months. As shown in Table~\ref{tab:online}, GeoGR-0.6B achieves consistent improvements across all evaluated scenarios. The average relative improvements are +2.91\% WINRATE, +2.02\% P\_CTR, +5.55\% PV\_CTR, and +1.68\% UV\_CTR. Notably, the Life Services scenario benefits the most from GeoGR, with improvements of +3.71\% WINRATE, +3.39\% P\_CTR, and +9.37\% PV\_CTR. This is likely because lifestyle queries (e.g., dining, entertainment, and local services) are more sensitive to spatio-temporal context and cross-category relationships, which GeoGR explicitly models through geo-aware SIDs and condition-aware generation. The Restaurant and Tourist Attraction scenarios also show steady gains, confirming the robustness of GeoGR across diverse LBS domains. Here, WINRATE is a composite engagement metric that weights strategic actions (clicks, navigation, bookings) by their business value, while P\_CTR, PV\_CTR, and UV\_CTR capture engagement at the item, page-view, and user levels, respectively. These results confirm that GeoGR effectively replaces traditional cascade architectures in large-scale industrial settings and delivers substantial business value through unified generative retrieval.

\begin{table}[t]
    \centering
    \caption{Relative improvement of GeoGR compared to the traditional cascade retrieval baseline of the online A/B test.}
    \scalebox{0.88}{
    \begin{tabular}{ccccc}
    \hline
    \hline
    Online metrics  & WINRATE &  P\_CTR &  PV\_CTR  & UV\_CTR\\
    \hline
    Restaurant          &+1.83\%   & +1.29\% &  +1.55\% & +0.82\%\\ 
    Life Services       &+3.71\%   & +3.39\% &  +9.37\% & +3.00\%\\ 
    Tourist Attraction  &+2.18\%   & +1.39\% &  +2.43\% & +1.21\%\\ 
    \hline
    Ave.Imp(0.6B, 3 months) & \textbf{+2.91\%}	&\textbf{+2.02\%} &	 \textbf{+5.55\%}	&\textbf{+1.68\%} \\ 

    \hline
    \hline
    \end{tabular}
    }
\label{tab:online}
\end{table}

\section{Conclusion}

In this work, we present GeoGR, a generative recommendation framework tailored for large-scale POI retrieval in navigation-based LBS scenarios. To construct geo-aware semantic POI SIDs, we first fine-tune the embedding encoder on sampled geo-constrained co-visited POI pairs, then apply RQ-Kmeans and an EM-style algorithm to generate and refine SIDs. To train a condition-aware generative recommendation model for industrial deployment, we further introduce a CPT and SFT training procedure. Extensive offline and online experiments demonstrate the effectiveness of each component and the superiority of GeoGR over strong baselines. GeoGR now serves millions of users daily in the AMAP App and delivers significant improvements in multiple metrics, confirming its practical value and scalability in large-scale LBS production environments.


\bibliographystyle{ACM-Reference-Format}
\bibliography{gr_bib}

\end{document}